# Converting vertical heat supply into horizontal motion for microtechnological pumping and autonomous waste heat recovery

Jan-Niklas Schäfer[1,*], Tillmann Carl[1,*], Kristin Kühl[1], Sonja Kiehren-Ehses[1], Jan Aurich[1], Georg von Freymann[3,4,5], Clarissa Schönecker[1,2,6]

[1]RPTU University Kaiserslautern-Landau, Kaiserslautern, Germany

[2]Darmstadt University of Applied Sciences, Darmstadt, Germany

[3]Department of Physics and Research Center OPTIMAS, RPTU University Kaiserslautern-Landau, Kaiserslautern, Germany

[4]Fraunhofer Institute for Industrial Mathematics ITWM, Kaiserslautern, Germany

[5]Institute of Photonic Systems and Technologies, Leibniz Universität Hannover, Hannover, Germany

[6]European University of Technology

*Equal contributions

## Abstract

The rapid advancement of high-performance computing infrastructure and its extended application produce an increasing amount of waste heat. This heat constitutes an unsustainable loss of energy as well as requires cooling solutions that transcend conventional thermal management. Here, we demonstrate a novel mechanism that converts vertical waste heat supply directly into horizontal fluid motion, enabling autonomous, self-powered pumping in microenvironments. Our approach is based on a concept that combines geometric symmetry breaking with heterogeneous thermal conductivities to induce local thermocapillary Marangoni flows. We provide an implementation of the concept as well as an experimental and numerical proof-of-concept, showing good agreement between the respective flow fields. The approach is scalable and operates under realistic areal heating conditions. It enables versatile pumping designs for microtechnological applications, lab-on-a-chip architectures, passive thermal management and heat-driven microfluidic systems.

## Introduction

Increasing the efficiency of power electronics is a key objective of modern energy systems. Despite considerable progress in minimising electrical losses, thermal energy dissipation remains an unavoidable by-product. In order to prevent overheating, it is necessary to incorporate cooling systems. Active cooling mechanisms, such as pumps or thermoelectric modules (e.g. Peltier elements), which are widely utilised for this purpose, require additional energy and add to the amount of energy losses. With upcoming technologies like Artificial Intelligence and Quantum Computing, several studies and reports predict an extraordinary increase of energy for corresponding data centres[1–3]. Also, the ongoing miniaturization of electronic devices has resulted in significant increases in packing densities and heat fluxes, making thermal management a key factor in their further development [4–9]. As microdevices still continue to scale down in size while increasing in power, there is an urgent need for compact, efficient, and integrable cooling solutions [4,5]. In summary, first, an increasing amount of waste heat will occur, and second, more advanced heat management technologies will be needed as appliances become smaller, energy density in systems continues to increase, and the number of systems to be cooled increases.

Ideally, the resultant waste heat would be used purposefully, for example to power or propel something. Thereby, the waste heat is not anymore only something to be removed, and for which extra energy might have to be spent. Rather, the available energy in the waste heat could be employed productively. To address this idea, this work presents a mechanism for converting vertical heat supply into horizontal motion, specifically fluid motion.

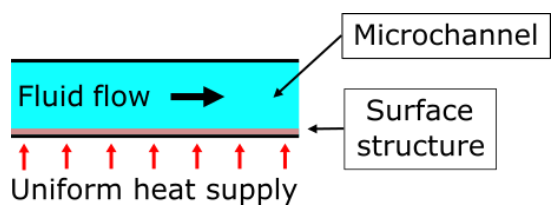

**Figure 1 | Operating Principle of the pump**. By means of a specifically structured bottom, vertical heat supply such as waste heat is converted autonomously into horizontal fluid motion.

The driving mechanism is based on the thermocapillary Marangoni effect. In general, thermocapillary forces occur at fluid-fluid interfaces if the surface tension σ is nonuniformly distributed. Due to the temperature-dependent nature of surface tension, this nonuniform distribution may be achieved by nonuniform heating of the fluid-fluid interface. The flow is then parallel to the interface.

Previously, several concepts for employing Marangoni forces for pumping have been explored both theoretically[10–12] and experimentally[13–17]. These however require the temperature gradient to be applied in the direction of fluid flow as it is typical for thermocapillary flows. Deviating concepts are rare and exist solely in abstract, theoretical form[11,18]. If we consider a typical source of waste heat, this longitudinal temperature gradient in the direction of flow past the surface is however often difficult to maintain in practice. Generally, we face a certain heated area while any motion can only take place laterally to this area.

Therefore, we present a new concept, its experimental implementation and its proof of principle for a novel kind of micropump that uses a uniform heat supply from the bottom of a channel and converts it to a lateral fluid motion (Fig. 1). The presented principle provides a scalable solution for self-powered microfluidic systems.

The concept is based on two features. The first feature is a superhydrophobic surface. Superhydrophobicity leads to the entrapment of air within the structures of the surface, such that the necessary fluid-fluid-interface for creating a thermocapillary Marangoni effect is provided. A superhydrophobic surface alone however would require a temperature gradient along the surface as in Baier et al[10] to create a Marangoni flow along the surface. Therefore, in order to create a component of the temperature gradient laterally to the surface, the second feature of the concept is the use of asymmetry, specifically a double asymmetry.

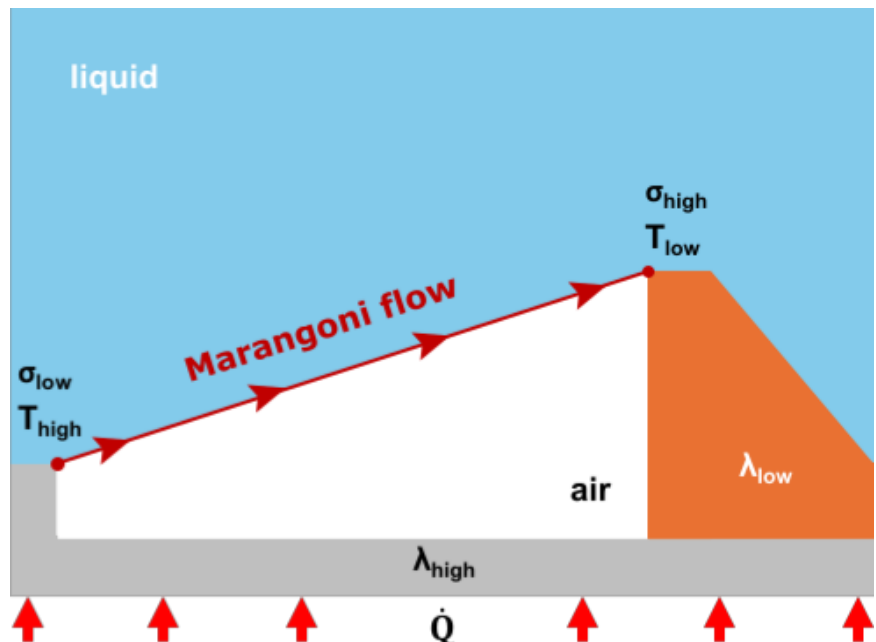

**Figure 2 | Schematic of the working principle.** The asymmetry in geometry and heat conductivity λ of the materials enables a component of the temperature gradient in the horizontal direction under heat supply from below. The resulting gradient of surface tension along the interface induces a net flow within the channel.

The double asymmetry consists of combining symmetry breaking in the geometry of the structures with the use of materials of different heat conductivities $\lambda$. Specifically, the base of the channel is composed of a material exhibiting a high thermal conductivity and thereby providing a good transmission of the heat. A material with a low heat conductivity and an asymmetric geometry is placed on top. Along the fluid-fluid interfaces, gradients in temperature $T$, and consequently, in surface tension will form and result in a flow along the interface in the direction of the lower temperature (see Fig. 2).

An arbitrary number of such segments can be connected in series to create a microchannel through which a liquid is pumped without the need for mechanical moving parts or other external energy, when powered by waste heat. In contrast to regular micropumps, our novel concept can then be classified as a "passive pump". While pumps are generally active, since they do require an external energy input, this pump can be considered self-powered, driven solely by autonomous waste heat recovery.

Besides the already mentioned advantage of a uniform heat supply, the concept possesses another advantage. Because of the small distances over which the temperature gradients are applied, the magnitude of the temperature differences can be comparatively small. In contrast, in systems that apply a temperature gradient along an interface[12–17], the temperature difference must be applied over a much greater distance.

# Results

## Implementation of the Concept

To demonstrate the functionality of the pumping concept, corresponding surfaces were manufactured (Fig. 3). The base structure was micromilled from aluminium, exhibiting a good thermal conductivity such that waste heat can be efficiently transported into the system. The structure on top was micro 3D printed with the photoresist so that both a material and a geometrical contrast is obtained. In order to enhance the pinning of the contact line at the upper edge of the printed structure, a small hook-like overhang structure was added. The whole surface was then rendered superhydrophobic by applying a nanoparticle spray.

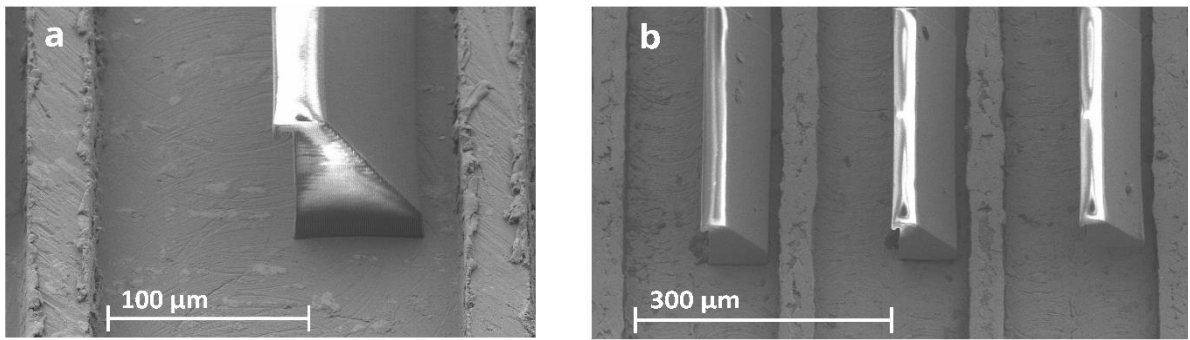

**Figure 3 | Manufactured aluminium and polymer microstructures on the sample block.** Images are captured by means of scanning electron microscopy. (a) Single pumping segment with the main groove's width of 100 µm, height of the aluminium bulge of 25 µm and height of the polymer of 90 µm. The total width of one pumping segment is 210 µm. (b) Several segments demonstrating the reoccurring pattern.

It was built into a microchannel system upside down, so that the region near the surface could be observed by laser scanning confocal microscopy. Throughout all experiments and up to the tested time frame of one day, the surface was reliably superhydrophobic, capturing air in the grooves as could be imaged by confocal microscopy (Fig. 4). Due to technical reasons the polymer had been printed at a short distance to aluminium bulge, so that a second small enclosure of air formed, as can be observed to the left of the aluminium bulge in Fig. 4.

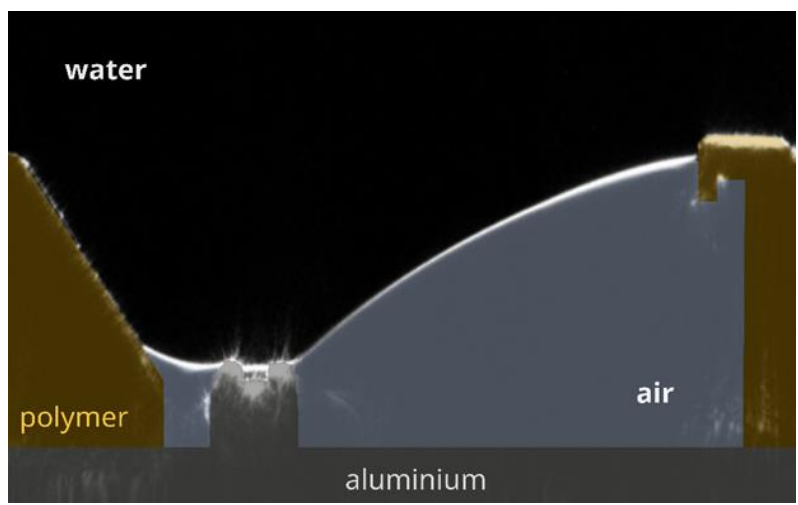

**Figure 4 | Confocal laser scanning image of a surface segment illustrating the formation of the water-air interfaces** (reflection mode, material colouring added to image). The small interface on the left is attributable to the printing process-constrained distance of the polymer to the aluminium bulge.

A comprehensive characterization of the interface's superhydrophobicity, including the apparent global contact angle and local contact angles of trapped large air inclusions, is provided in the Supplementary Information S1.

## Proof of concept

Pumping experiments involving confocal laser scanning and laser microparticle tracking velocimetry (µPTV) were conducted. The backside of the aluminium block was heated and microscopy images taken in the x-z plane. To visualize the flow pattern, particle tracking velocimetry was performed using fluorescent tracer particles (size 2 µm). Image sequences were analysed with custom-written Python routines based on the Trackpy library for particle tracking and StarDist neural network AI for particle recognition.

Furthermore, steady-state numerical simulations of the coupled flow and heat transfer were performed using the commercial software COMSOL Multiphysics. The geometry of the interfaces was imported from the microscopy images.

Fig. 5 displays an overlay of the experimentally measured and simulated velocity fields within the microchannel. The continuous streamlines correspond to the simulated flow field, while the red arrows indicate the experimentally measured velocity vectors. Both were normalised to the maximum flow velocity.

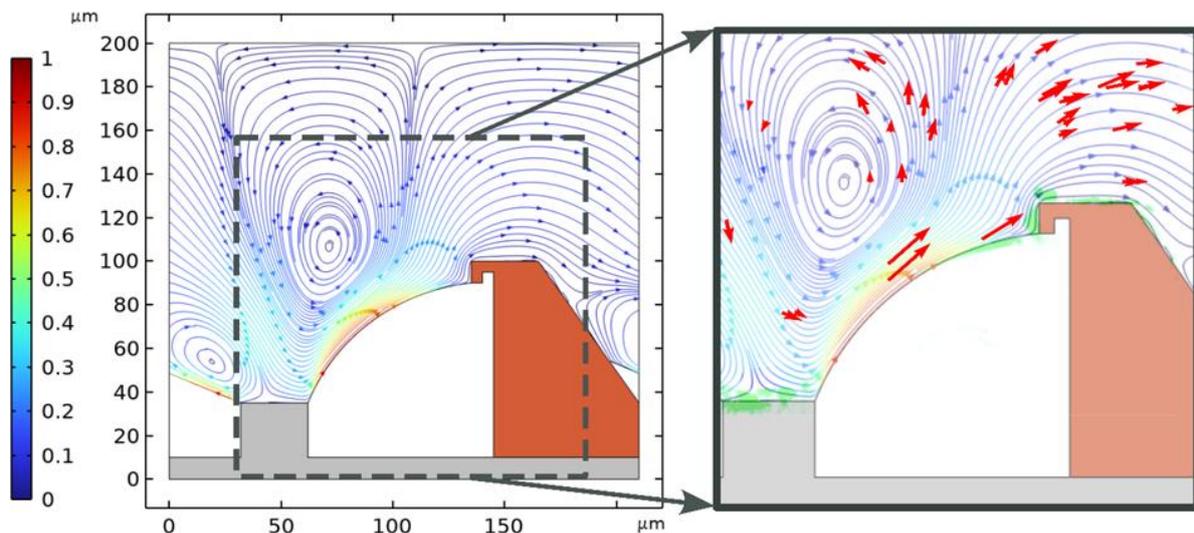

**Figure 5 | Experimental and numerical flow field during autonomous pumping process.** Left: Numerically computed flow field in the channel (colouring: velocity magnitude). Right: Detailed view of the highlighted cutout region. Experimentally measured red velocity vectors and confocal interfacial image are superimposed (green coloured couture: reflection channel microscopy).

Evidently, the experimental and simulation approaches result in a consistent flow pattern. The highest velocity clearly is directly at the water-air interface where the thermocapillary forces act. These forces drive a bulk flow along the channel and create a net flow in horizontal direction that is determinable by both numerical and experimental means. Above the air-water interface and at about the widest cross section, a large central vortex is formed. Its shape and spatial extent are somewhat influenced by the disparity in height and width of the individual microstructures, as suggested by simulations with varying geometric configurations.

While the practically applied temperature difference is located between the outer sides of the microchannel, the physically relevant temperature difference is between the two microstructures. It evolves on the one hand through heat conduction through the various solid materials of the channel and on the other hand through the heat transport in the established flow.

In the experiments, the actual temperature difference across the channel cannot be assessed, therefore no direct comparison of the absolute velocity values is possible. Furthermore, the experimental heating process was instationary. At a point in time at the beginning of the heating process that corresponds to Fig. 5 right, an average velocity above the polymer structures of about 50 µm/s can be estimated from the experimental data. The low particle density employed for particle tracking as well as the limited frame rate of the microscope impede a statistical evaluation of faster velocities at a later point in time.

The numerical simulations cover a broader range than the actual experiments, enabling a systematic investigation of the temperature difference dependence of the achieved flow (see Fig. 6). Under a temperature difference of $\Delta T$ = 0.5K applied between the top and bottom boundaries (as in Fig. 5 left), the simulations predict the pump to generate a mean flow velocity of 425 µm/s, measured above the polymer structure.

In practice, this value may be reduced by the presence of contaminations at the air-water interface, e.g. by the natural occurrence of impurities or specifically in the present experiment by tracer particle accumulation at the interface. Such contaminations are known to reduce the effective slip length[19,20], but cannot be quantified and therefore not be simulated. Previous experiments reported a factor of 2 difference in the Marangoni velocity at a superhydrophobic surface between experiments and an idealized simulation[13].

Overall, given the short distances of both the microchannel height as well as the height difference between the two microstructures, the absolute values of the temperature differences in order to obtain a flow, are also very low.

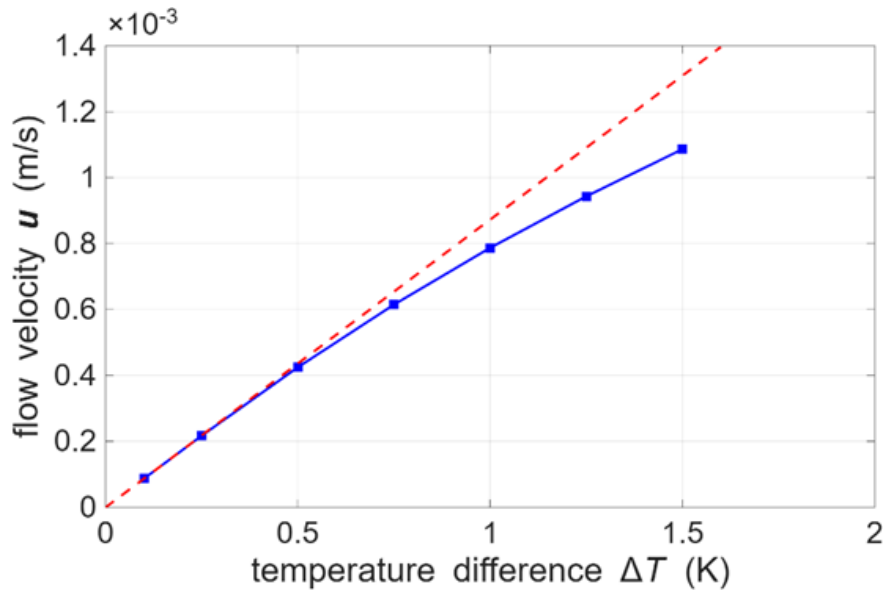

**Figure 6 | Flow velocity as a function of the applied temperature difference ΔT across the channel height.** The given flow velocity is the average velocity in the direction parallel to the surface, evaluated in the cross section above the polymer structure. The red dashed curve represents a linear slope overlapping with the initial onset of the blue interpolation curve of numerical simulated data points.

Another observation can be drawn from Fig. 5. While the experimental observation is limited to a region close to the air-water interface due to optical reasons, the numerical calculations can also illustrate the velocity field beyond this region. Due to manufacturing constraints that led to the polymer being printed at a short distance from the aluminium bulge, a second small enclosure of air occurred in the experiments. From the numerical calculations, it can be observed that the overall horizontal flow is weakened by a small vortex at this secondary interface. The second vortex can be attributed to the occurrence of Marangoni stresses at this second interface, which act in opposition to the primary flow direction. Consequently, it is to be expected that with advanced manufacturing conditions, the second vortex could be removed, and the net flow could be increased. This is confirmed by numerical simulations. Specifically, in the simplest scenario, changing the boundary condition of the secondary interface to a regular no-slip wall, i.e. pretending this interface to be solid, already results in a remarkable 18-20% increase in the simulated volumetric flow rate, dependent on the applied temperature gradient. It is expected that more optimized surface geometries, e.g. with a larger ratio of air-water interface area to water-solid interface area, will result in even higher velocities and flow rates.

## Conclusions

A concept for converting vertical heat supply into horizontal fluid motion has been presented. It was successfully implemented and validated though both experimental measurements and numerical simulation. With its ability to autonomously recover waste heat, the concept marks a starting point for sustainable pumping in microgeometries.

Compared to other thermocapillary pumping principles, the principle has several advantages. It works with a uniform heat supply from the side of the channel, that it is not required to apply the driving temperature difference along the channel. A uniform areal heat supply corresponds to practically occurring scenarios, e.g. in microelectronic devices. Due to the small distances involved in the asymmetry of the surface structure, the pumping furthermore works with comparatively small temperature gradients enabling a potentially use in various microfluidic devices.

An arbitrary number of pumping elements can be placed behind each other and thus constitutes a modular system. Since the temperature difference is not applied along the channel but across it, the modular principle also enables new versatile microfluidic designs where the flow takes curves or takes more complicated routes on a chip that have not been possible before.

By optimising various influencing parameters, such as the geometry of the structures, the pumping efficiency can be increased, and higher fluid velocities will be achieved.

The presented concept offers possibilities for a wide range of new applications. In future, this mechanism might for instance serve to be implemented in lab-on-a-chip applications or as a sustainable cooling mechanism capable of dissipating waste heat without the use of external energy. Additionally, the concept offers potential for solar-driven applications in which the heat is provided via a radiation-absorbing coating.

# Methods

## Materials

Spherical red fluorescent polystyrene microparticles of diameter 2 µm were purchased from micromod Partikeltechnologie GmbH (Rostock, Germany), precision cover glass No. 1.5H/ 170 µm thickness was purchased from Paul Marienfeld GmbH & Co. KG (Lauda-Königshofen, Germany), heat resistant glue WoldoClean Professional SUPERGLUE universal was purchased from Woldoshop GmbH (Norderstedt, Germany), photoresist for printed structures IP-S photoresist is purchased from Nanoscribe GmbH & Co. KG (Karlsruhe, Germany), superhydrophobic nanoparticle spray Glaco Mirror Coat Zero was purchased from Soft99 (2-6-5 Tanimachi, Chuo-ku, Osaka, Japan).

## Fabrication of the surface

### Micromilling Aluminium block

An aluminium block (AlCuMgSi - 10 mm x 10 mm x 4 mm) was microstructured on an Ultra-Precision-Milling Machine (MMC 900H – LT Ultra). First, the surface was facemilled with an end mill (diameter 3 mm). The desired microstructure was then micro milled into the surface using a micro end mill with a diameter of 100 µm. A spindle speed of n =

30,000 rpm, a depth of cut of ap = 5 µm and a feed per tooth of fz = 1 µm were used. The slots were produced according to Fig. 3. Slot widths of 180 µm, slot depths of 25 µm and bar widths of 30 µm were determined.

**Printing of 3D-structures**

To complete the pumping structures, the polymer part was printed via Direct Laser Writing. Direct Laser Writing is a versatile technique based on a two-photon lithography process that enables printing of polymer structures with feature sizes down to the 100 nm range [21,22]. A pulsed laser is focused tightly into a liquid photoresist containing monomers as well as a photoinitiator, which is excited via two photon absorption to ensure a very small volume of excitation. As a result, the photoinitiator molecules in this volume fall apart to radicals, which bind to the monomers thus starting a photopolymerization process creating a polymerised volume. By moving the focus in the liquid resist, detailed three-dimensional structures can be generated. In the final step, the remaining liquid photoresist is washed away by a solvent, leaving the printed structure on the substrate [22,23].

For this work, a Photonic Professional GT+ (Nanoscribe GmbH & Co. KG) was used in combination with a commercial photoresist (IP-S, Nanoscribe GmbH & Co. KG). The corresponding printing software called Nanowrite offers the possibility for a coordinate transformation using markers that can be placed on the printing substrate based on a camera image. Thus, the printed polymer structures were aligned to the slots on the micromilled aluminum block and foil prepared as described in the previous section. The printed structure expands throughout each slot in length, consisting of 300 µm long pieces for technical reasons, with a width of 65 µm and height of 90 µm. The hook on the upper corner of the structure features a size of 5 µm in width. SEM pictures of the printed structure aligned in the micromilled slots on the aluminum block are shown in Fig. 3.

**Coating of the microstructures**

First, coating tests were performed to determine the optimal coating thickness for achieving the desired water–air interface when the channel is filled with water. After coating, the water–air interfaces were characterised in aqueous environment using laser scanning microscopy (LSM) with the Leica Stellaris 8 inverted confocal microscope equipped with a water-immersion objective (HC PL APO 63×/NA 1.20).

For the final coating procedure, the aluminium block was placed flat on a laboratory table with the microstructures facing upward. The superhydrophobic coating spray was thoroughly shaken, and the nozzle was cleaned with a laboratory wipe to remove any contamination. The spray was pre-shot onto a wipe to prefill the nozzle and ensure consistent application. Excess material at the nozzle outlet was removed with a wipe. The entire surface was then sprayed once across from left to right at a distance of approximately 10 cm, producing a uniform thin liquid layer. Small amounts of excess coating at the edges of the aluminium block were removed immediately. The sample was placed in a large, closed box to allow the liquid to dry. Nanoparticles remaining on the surface formed a superhydrophobic coating.

## Experimental setup

### Preparation of channel

The aluminium block with the superhydrophobic microstructures on its bottom side was glued onto a cover glass (Fig. 7). Two spacers separated this cover glass from another cover glass in order to later confine a water layer between the bottom cover glass and the aluminium block. To retain the water, a reservoir was constructed on the bottom cover glass. The geometry was chosen such that the block contains many pumping elements along its length, while the width of a single element was much larger than the height of the confined water layer. This geometrical design ensured that the flow in the confined water at the centre of the aluminium block could be measured avoiding boundary effects of the sample. The spacers had a low thermal conductivity (estimated to 0.12 W m$^{-1}$ K$^{-1}$) compared to aluminium (236 W m$^{-1}$ K$^{-1}$) and glass (0.76 W m$^{-1}$ K$^{-1}$), and thereby minimised heat transfer from the aluminium block into the water pool. In addition, the spacers were positioned at the edges of the cover glass, far from the aluminium block, creating long lateral heat paths.

For filling the channel with water, the aluminium block carrying the superhydrophobic microstructures (Fig. 7) was oriented aluminium block facing upwards. Approximately 90 µL of a suspension of 2 µm sized fluorescent polystyrene particles (concentration of 1.7 billion particles per ml) was pipetted onto the interface in three aliquots of 30 µL each. This procedure avoided immediate displacement of the entire drop along the superhydrophobic surface due to its high hydrophobicity. When the bottom cover glass with the glued spacers was subsequently placed on top, the three droplets merged into one continuous liquid volume within the channel spacing. To ensure complete filling, the sample was carefully inverted, and additional solution was introduced from the side with a pipette. Finally, the surrounding pool was filled with ≈60 µL of the same solution, resulting in a shallow liquid reservoir. Due to capillary effects, the confined liquid formed slightly curved menisci at the channel edges, smoothly connecting to the reservoir. The assembly was fixed at both ends with adhesive tape spanning the top and bottom cover glasses.

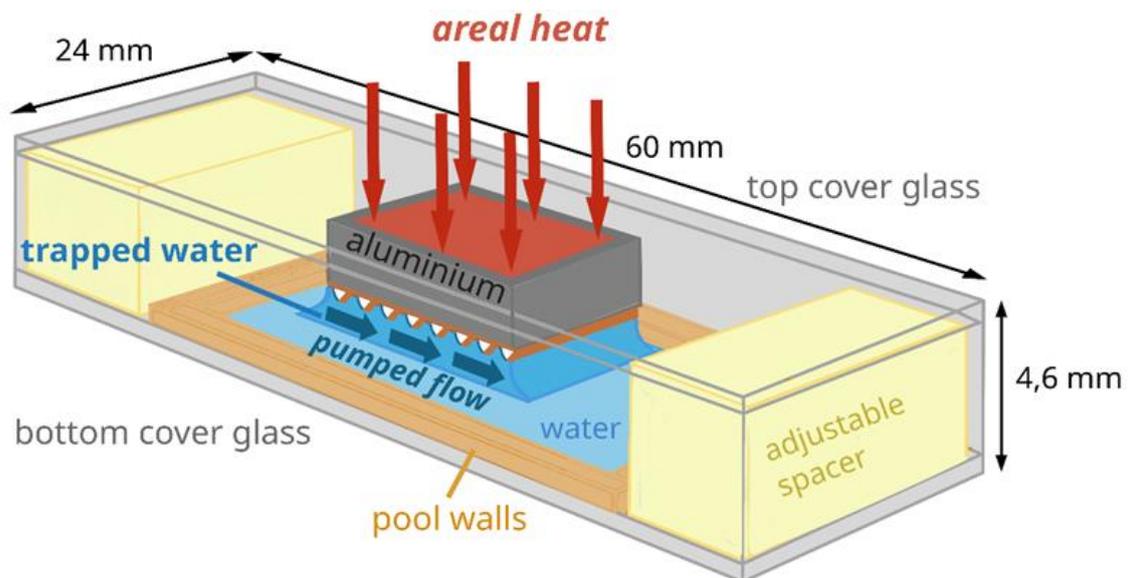

**Figure 7 | Experimental sample Marangoni pump in a microchannel with open side walls.** An areal heat is applied on the aluminium block, which drives a flow at the superhydrophobic pumping interface building up a pumping mechanism through the channel.

The pumped flow was generated by heating the entire top surface of the aluminium block, inducing a Marangoni flow along the superhydrophobic interface. A butane stick lighter was used as a heat source. Using mainly radiative heating minimised mechanical weight load on the sample. Heating from the top prevented buoyancy-driven vortices, since the heated surface was on top and the cold surface was at the bottom.

The pump flow dynamics were characterised using microparticle tracking velocimetry (micro-PTV) with the Leica Stellaris 8 confocal laser scanning microscope and 2 µm sized fluorescent particles (Fig. 8). Measurements were performed in the confined water layer at the centre of the channel to avoid side effects from the open channel edges, providing conditions comparable to a fully closed microchannel. Micro-PTV measurements were conducted by scanning the region of interest in the x-z plane with a pulsed white light laser with an emission wavelength of 552 nm. Simultaneously, the scanning geometries were controlled in reflection mode using the 440 nm emission of this laser, managed via an acousto-optical beam splitter. A comparatively low optical resolution of 64 x 64 pixel for an image size of 145 x 145 µm was chosen to achieve sufficiently high frame rates of 4,05 frames/s with a pixel dwell time of 6,59 µs to capture the pumping dynamics around the Marangoni interface for analysis in Python.

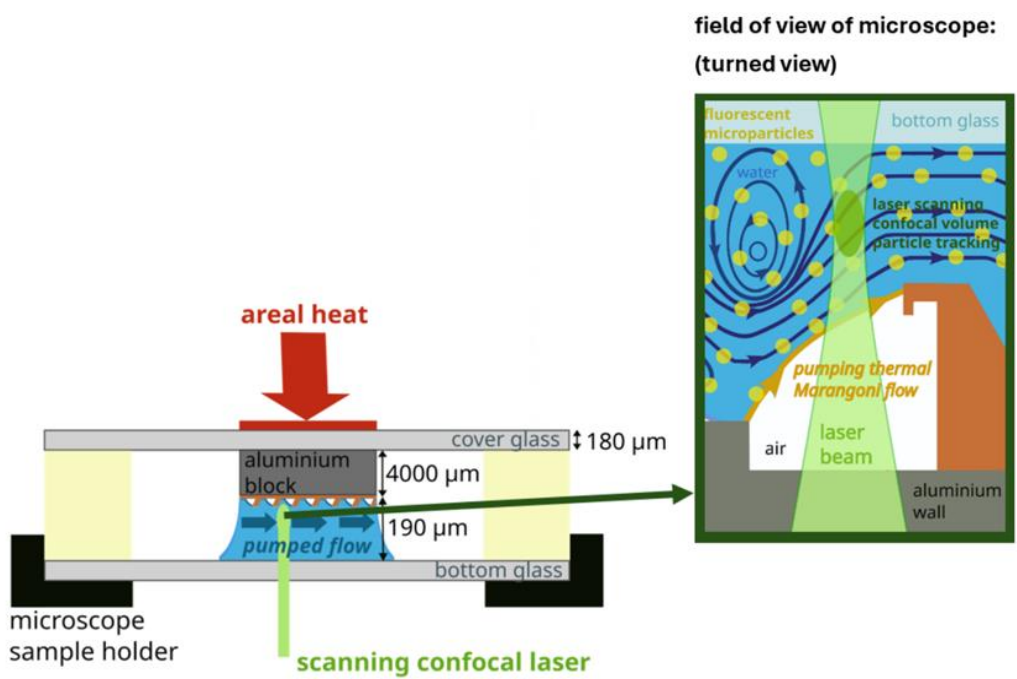

**Figure 8 | Experimental setup for confocal laser microparticle tracking**. On the left figure side is the schematic sample of the microchannel in the microscope holder during microparticle tracking with a scanning confocal laser. On the right side, there is a schematic of

the confocal scanned field of view, biggest possible image size, captured with a water immersion objective.

**Data analysis**

Particle tracking data were analysed using custom Python 3 scripts (version 3.12.3, 64 Bit) for particle tracking analysis with the software package Trackpy (version 0.6.2rc1) and the deep-learning-based particle recognition tool StarDist (version 0.9.1). Trackpy provides algorithms to link particles across frames to form trajectories, enabling the calculation of properties such as velocity and diffusion trajectories. This approach allows precise analysis of low velocities in microsystems where diffusion plays a significant role and includes correction for thermal focus drifts during laser scanning. StarDist detects and segments approximately round or star-convex objects by representing each particle as a star-convex polygon, connecting points along the boundary with beams radiating from the particles centre. When an object changes shape between frames, StarDist matches it to the most similar object in a defined neighbourhood, allowing accurate tracking of overlapping or touching particles, which is important for high particle concentrations or particle agglomeration. For evaluating the flow velocities in Python, the pixel size of one frame was transformed into meters and divided by the corresponding frame interval time. To validate the accuracy of the velocity evaluation performed with Trackpy and StarDist, particle velocities were additionally determined manually in ImageJ. For this, the displacement of individual particles between consecutive frames was measured using the distance tool and divided by the corresponding frame interval time. StarDist was already pre-trained on images of small round particles to provide a reference for recognition, as the recorded particles often appeared as asymmetric ellipses due to irregular illumination caused by diffraction from sedimented particles and time-dependent scanning in the confocal microscope. For a more detailed explanation of the used Python script see the supplementary information S2.

**Simulations**

The numerical model for three-dimensional flow field calculation was built and solved with the software COMSOL Multiphysics 6.1. The simulation domain comprises one single segment of the pump with periodic boundary conditions for the in- and outlet. It includes the domains for the aluminium bottom, the printed polymer structures, the air pockets and the water. The model couples Stokes equation and mass conservation solved in the water domain with heat transfer in the entire simulation domain. The coupling is realized by the implementation of a temperature-dependent boundary stress condition at both water-air interfaces including the thermocapillary stress $\frac{d\sigma}{dx} = \frac{d\sigma}{dT}\frac{dT}{dx}$, with $\frac{d\sigma}{dT}$ being assumed constant. Dirichlet temperature boundary conditions are applied at the channel walls. The upper wall of the channel is maintained at a constant temperature of $T_{bottom}$ =293.15 K, while the lower boundary is prescribed a higher temperature of $T_{top}$

=293.15 K+ΔT. Within the water domain, heat transfer occurs through both conduction and convection.

$$\rho\, c_p\, \boldsymbol{u} \cdot \nabla T + \nabla \cdot (-\lambda\, \nabla T) = 0$$

In all other domains, heat transfer is modelled exclusively by thermal conduction. The temperature derivative of the surface tension $\frac{d\sigma}{dT}$ is set to be -0.00015 N/(m*K). The printed microstructure (hardened photoresist) was parameterised with a heat conductivity $\lambda$ of 0.2 W/(m*K), density ρ of 1.19 g/cm³ and a heat capacity $c_p$ of 1.6 J/(g*K). The properties of air, water and aluminium are obtained from the Comsol material library.

# Data availability

The data supporting the findings of this study are available within the paper and its Supplements. Data is stored as Supplementary Data on Figshare or may be requested from the corresponding author upon reasonable request.

# Code availability

The python script necessary to evaluate the paper's results is available within the paper's supplementary data.

# References

1. Koot, M. & Wijnhoven, F. Usage impact on data center electricity needs: A system dynamic forecasting model. *Applied Energy* **291,** 116798; 10.1016/j.apenergy.2021.116798 (2021).
2. Electricity 2024. International Energy Agency (IEA), 2024.
3. Mytton, D. & Ashtine, M. Sources of data center energy estimates: A comprehensive review. *Joule* **6,** 2032–2056; 10.1016/j.joule.2022.07.011 (2022).
4. Rangarajan, S., Schiffres, S. N. & Sammakia, B. A Review of Recent Developments in "On-Chip" Embedded Cooling Technologies for Heterogeneous Integrated Applications. *Engineering* **26,** 185–197; 10.1016/j.eng.2022.10.019 (2023).
5. Fazeli, K. & Vafai, K. Analysis of optimized combined microchannel and heat pipes for electronics cooling. *International Journal of Heat and Mass Transfer* **219,** 124842; 10.1016/j.ijheatmasstransfer.2023.124842 (2024).
6. Dhumal, A. R., Kulkarni, A. P. & Ambhore, N. H. A comprehensive review on thermal management of electronic devices. *J. Eng. Appl. Sci.* **70**; 10.1186/s44147-023-00309-2 (2023).
7. Zahid, I., Farhan, M., Farooq, M., Asim, M. & Imran, M. Experimental investigation for thermal performance enhancement of various heat sinks using Al2O3 NePCM for cooling of electronic devices. *Case Studies in Thermal Engineering* **41,** 102553; 10.1016/j.csite.2022.102553 (2023).

8. Rakshith, B. L. *et al*. Cooling of high heat flux miniaturized electronic devices using thermal ground plane: An overview. *Renewable and Sustainable Energy Reviews* **170,** 112956; 10.1016/j.rser.2022.112956 (2022).

9. Sun, B. & Huang, X. Seeking advanced thermal management for stretchable electronics. *npj Flex Electron* **5**; 10.1038/s41528-021-00109-9 (2021).

10. Baier, T., Steffes, C. & Hardt, S. Thermocapillary flow on superhydrophobic surfaces. *Physical review. E, Statistical, nonlinear, and soft matter physics* **82,** 37301; 10.1103/PhysRevE.82.037301 (2010).

11. Crowdy, D., Mayer, M. & Hodes, M. Asymmetric thermocapillarity-based pump: Concept and exactly solved model. *Phys. Rev. Fluids* **8**; 10.1103/PhysRevFluids.8.094201 (2023).

12. Yariv, E. Thermocapillary flow between longitudinally grooved superhydrophobic surfaces. *J. Fluid Mech*. **855,** 574–594; 10.1017/jfm.2018.686 (2018).

13. Gao, A., Butt, H.-J., Steffen, W. & Schönecker, C. Optical Manipulation of Liquids by Thermal Marangoni Flow along the Air-Water Interfaces of a Superhydrophobic Surface. *Langmuir : the ACS journal of surfaces and colloids* **37,** 8677–8686; 10.1021/acs.langmuir.1c00539 (2021).

14. Amador, G. J. *et al*. Temperature Gradients Drive Bulk Flow Within Microchannel Lined by Fluid-Fluid Interfaces. *Small (Weinheim an der Bergstrasse, Germany)* **15,** e1900472; 10.1002/smll.201900472 (2019).

15. Sammarco, T. S. & Burns, M. A. Thermocapillary pumping of discrete drops in microfabricated analysis devices. *AIChE Journal* **45,** 350–366; 10.1002/aic.690450215 (1999).

16. Debar, M. J. & Liepmann, D. Fabrication and performance testing of a steady thermocapillary pump with no moving parts. In *Technical Digest. MEMS 2002 IEEE International Conference. Fifteenth IEEE International Conference on Micro Electro Mechanical Systems (Cat. No.02CH37266)* (IEEE2002), pp. 109–112.

17. Frumkin, V., Gommed, K. & Bercovici, M. Dipolar thermocapillary motor and swimmer. *Phys. Rev. Fluids* **4**; 10.1103/PhysRevFluids.4.074002 (2019).

18. Mayer, M. D., Kirk, T. L., Hodes, M. & Crowdy, D. Mechanical power from thermocapillarity on superhydrophobic surfaces. *J. Fluid Mech*. **1009**; 10.1017/jfm.2025.188 (2025).

19. Schäffel, D., Koynov, K., Vollmer, D., Butt, H.-J & Schönecker, C. Local Flow Field and Slip Length of Superhydrophobic Surfaces. *Physical review letters* **116,** 134501; 10.1103/PhysRevLett.116.134501 (2016).

20. Bolognesi, G., Cottin-Bizonne, C. & Pirat, C. Evidence of slippage breakdown for a superhydrophobic microchannel. *Physics of Fluids* **26**; 10.1063/1.4892082 (2014).

21. Maruo, S., Nakamura, O. & Kawata, S. Three-dimensional microfabrication with two-photon-absorbed photopolymerization. *Optics letters* **22,** 132–134; 10.1364/OL.22.000132 (1997).

22. Skliutas, E. *et al*. Multiphoton 3D lithography. *Nat Rev Methods Primers* **5**; 10.1038/s43586-025-00386-y (2025).

23. Young, O. M., Xu, X., Sarker, S. & Sochol, R. D. Direct laser writing-enabled 3D printing strategies for microfluidic applications. *Lab on a chip* **24,** 2371–2396; 10.1039/D3LC00743J (2024).

## Acknowledgements

We kindly acknowledge the funding by the Deutsche Forschungsgemeinschaft (DFG, German Research Foundation) - project number 501107071. We further acknowledge the support of the Leica Stellaris 8 confocal microscope, funded by the Deutsche Forschungsgemeinschaft - project number 503863443.

## Author Contributions

J.-N.S. designed and performed the experiments, analysed the results including the coding of the Python script for data analysis. T.C. performed the numerical simulations and analysed the results. T.C., J.-N.S. and C.S. discussed the results, including experiments and simulations. All authors discussed the fabrication of the sample with micromilling and 3-D-printing. K.K. did the 3-D-printing of the polymer microstructure and SEM-imaging of the sample. S. K.-E. did the micromilling of the aluminium microstructure. J.-N.S. did confocal imaging and contact angle measurements. C.S. directed the project and supervised data analysis. J.-N.S., T.C., C.S., K.K. and S.K.-E. contributed to writing the paper and revising. J.A. und G.v.F. contributed to the discussion and reviewing the manuscript.

### Corresponding author

Correspondence and requests for materials should be addressed to Clarissa Schönecker (Email: clarissa.schoenecker@h-da.de)

## Competing Interests

The authors declare no conflict of interest.

# Supplementary Information

## S1: Apparent and local contact angle characterisation

### a. apparent contact angle: water drop on the pump's superhydrophobic interface

Apparent contact angles on the pump's superhydrophobic interface were measured using a Krüss DSA 30 drop-shape analyser. A 5 µl droplet of ultrapure water was deposited onto the interface by first forming the droplet at a needle tip, slowly lowering it until contact of the drop with the interface was established and then the tip was carefully automatically removed. To prevent electrostatic charging of the droplet, the needle was grounded using a copper wire. Contact angle experiments were conducted in a climate-controlled chamber at a temperature of 20 °C and a relative humidity of 35%. To ensure consistent surface conditions and the absence of a pre-wetting layer, all droplets were deposited onto a dry interface. Data analysis was performed using the Krüss Advance software, typically employing the ellipse fitting method for contact angle determination.

Due to the structural anisotropy of the superhydrophobic interface, contact angles were measured both parallel and transverse to the printed polymer. The measured static contact angles are 138° transverse to the polymer and 143° parallel to it, confirming the pronounced superhydrophobic behaviour of the interface described in the main paper text.

### b. Local contact angle: large air pocket in the superhydrophobic structure

Local contact angles at a single element were determined manually for a significant series of frames of the laser scanning recordings, captured during pump operation, using the angle measurement tool in ImageJ. Figure A1 left shows the schematically image of a cutout area of the superhydrophobic interface of the pump. The local contact angle of the air at the bottom pinning area between the water-air-interface and the aluminium bulge interface is defined as $α_{aluminium}$ and at the top pinning area the contact angle between the water-air-interface and the polymer hook´s interface is defined as $α_{polymer}$. A representative confocal image frame for this system is shown in Figure A1 (right), highlighting a large air pocket within the superhydrophobic structure. The measured angles $α_{aluminium}$ = 63° and $α_{polymer}$ = 84° are marked in this Figure. The contact angles on both sides of the pinned air pocket remain remarkably constant throughout the pumping duration. The results demonstrate that the interface maintains its stability and equilibrium morphology during operation. This robust Cassie state, which is essential for a sustainable pumping mechanism, confirms the efficacy of both the microstructural geometric design and the superhydrophobic coating.

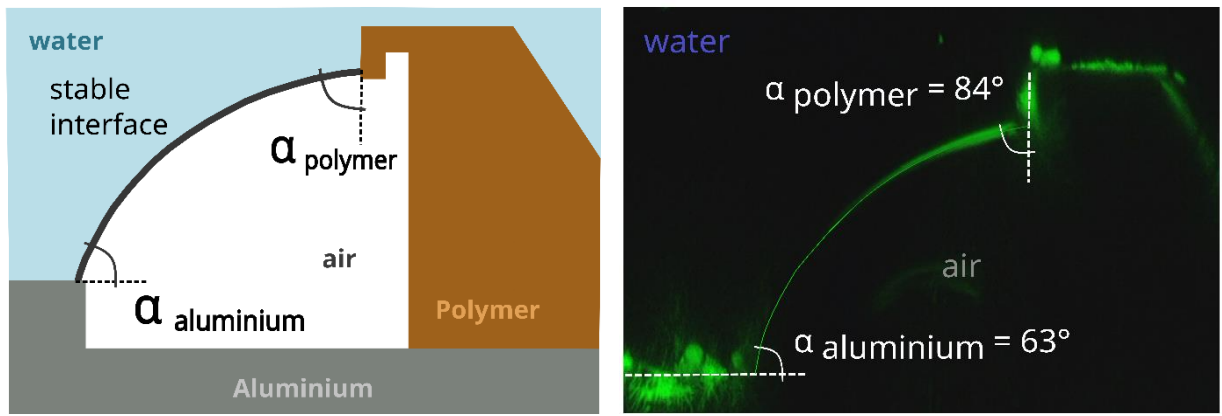

**Fig. A1 | Characterisation of pinning areas of air pocket inclusion.** Left: schematic representation of a cutout area of the pump´s superhydrophobic interface with the trapped large air pocket in the interfacial structure. The constant angles α$_{aluminium}$ and α$_{polymer}$ are the local contact angles of the stable water-air interface with the polymer hook's interface and aluminium's interface in the pinning area of this surface. Right: confocal laser scanning image of a cutout area of the superhydrophobic interface of the pump. To enhance visual clarity, the water-air interface is delineated with a thin green line. This overlay was applied to ensure the interface remains clearly discernible outside of the dark high-contrast environment of the microscope room, particularly in standard viewing conditions.

## S2: python video preprocessing and particle tracking workflow

Videos of fluorescence microscopy experiments were pre-processed and analysed using a custom Python pipeline combining open-source image analysis libraries (pims, trackpy, StarDist, csbdeep, and moviepy). Individual video frames were first extracted and converted into grayscale images by isolating the green fluorescence channel, which provided the highest signal-to-noise ratio for subsequent recognition accuracy. This preprocessing step was implemented using the *pims* library as a pipeline decorator, allowing automated channel selection for each frame during video loading. To obtain short video segments in which the flow velocity remains effectively constant, long microscopy recordings were automatically divided into smaller video segments. This was accomplished by iterating through all frames using *pims* and saving each subset as a separate .avi file via *moviepy*. Each segment was stored in a dedicated output directory to ensure compatibility with the subsequent object detection and tracking workflow.

For object segmentation, individual frames were processed using a pretrained *StarDist2D* model, which applies a deep-learning-based star-convex shape representation to delineate fluorescent tracer particles. The segmented regions were converted to labelled masks using *skimage.measure*, and particle centroids were extracted as input for trajectory linking. Particle tracking was performed using the *trackpy* library, which links detected particles across frames based on spatial proximity and expected displacement. The resulting trajectories were exported as *pandas* DataFrames for quantitative analysis.

Derived quantities such as mean displacement, velocity, and trajectory length were computed and statistically evaluated across multiple video segments.

Visualization of the flow fields was achieved using *matplotlib*, where individual vector arrows overlaid on the raw or segmented images. This modular workflow enabled reproducible, script-based preprocessing, segmentation, and quantitative tracking of fluorescent tracers in microfluid-systems videos, ensuring full transparency and flexibility for subsequent data analysis. All intermediate steps (videos, masks, data) are saved so that the results can be tracked at any time and parameters can be optimized iteratively.

Manual selective spot-checks were performed by visually tracking individual particles frame by frame. This step was necessary, in case the automated detection by StarDist produced mismatches, for instance when a particle from another focal plane temporarily entered the 2D plane and appeared similar to an existing moving particle whose brightness and shape changed slightly between frames. Manual verification and correction were also required due to inhomogeneous illumination conditions such as through strong reflections from the aluminium surface or light of partial sedimented particles on the cover glass. As mentioned in the main paper text, the velocities estimated of Python were checked manually measuring the displacement of individual particles between consecutive frames using the distance tool in ImageJ and dividing this distance by the corresponding frame interval time.